\documentclass[11pt]{article}
\usepackage{amsfonts,amsmath,amssymb,amsfonts,graphicx,hyperref}
\begin{document}

\title{Adaptive Wave Models for Option Pricing Evolution:\\ Nonlinear and Quantum Schr\"{o}dinger Approaches}
\author{Vladimir G. Ivancevic \\
{\small Defence Science \& Technology Organisation, Australia}}
\date{}
\maketitle

\begin{abstract}
Adaptive wave model for financial option pricing is proposed, as a high-complexity alternative to the standard Black--Scholes model. The new option-pricing model, representing a controlled Brownian motion, includes two wave-type approaches: nonlinear and quantum, both based on (adaptive form of) the Schr\"{o}dinger equation. The nonlinear approach comes in two flavors: (i) for the case of constant volatility, it is defined by a single adaptive nonlinear Schr\"{o}dinger (NLS) equation, while for the case of stochastic volatility, it is defined by an adaptive Manakov system of two coupled NLS equations. The linear quantum approach is defined in terms of de Broglie's plane waves and free-particle Schr\"{o}dinger equation. In this approach, financial variables have quantum-mechanical interpretation and satisfy the Heisenberg-type uncertainty
relations. Both models are capable of successful fitting of the Black--Scholes data, as well as defining Greeks.\\

adaptive Manakov system, quantum-mechanical option pricing, market-heat potential\\

\end{abstract}


\newpage

\section{Introduction}

Recall that the celebrated Black--Scholes partial differential equation
(PDE) describes the time--evolution of the market value of a \textit{stock
option} \cite{BS,Merton}. Formally, for a function $u=u(t,s)$ defined on the
domain $0\leq s<\infty ,~0\leq t\leq T$ and describing the market value of a
stock option with the stock (asset) price $s$, the \emph{Black--Scholes PDE}
can be written (using the physicist notation: $\partial _{z}u=\partial
u/\partial z$) as a diffusion--type equation:
\begin{equation}
\partial _{t}u=-\frac{1}{2}(\sigma s)^{2}\,\partial _{ss}u-rs\,\partial
_{s}u+ru,  \label{BS}
\end{equation}
where $\sigma >0$ is the standard deviation, or \emph{volatility} of $s$, $r$
is the short--term prevailing continuously--compounded risk--free interest
rate, and $T>0$ is the time to maturity of the stock option. In this
formulation it is assumed that the \emph{underlying} (typically the stock)
follows a \emph{geometric Brownian motion} with `drift' $\mu $ and
volatility $\sigma $, given by the stochastic differential equation (SDE)
\cite{Osborne}
\begin{equation}
ds(t)=\mu s(t)dt+\sigma s(t)dW(t),  \label{gbm}
\end{equation}
where $W$ is the standard Wiener process. The Black-Scholes PDE (\ref{BS})
is usually derived from SDEs describing the geometric Brownian motion (\ref%
{gbm}), with the stock-price solution given by:
\begin{equation*}
s(t)=s(0)\,\mathrm{e}^{(\mu -\frac{1}{2}\sigma ^{2})t+\sigma W(t)}.
\end{equation*}
In mathematical finance, derivation is usually performed using It\^{o} lemma
\cite{Ito} (assuming that the underlying asset obeys the It\^{o} SDE), while
in physics it is performed using Stratonovich interpretation \cite{Perello,Gardiner} (assuming that
the underlying asset obeys the Stratonovich SDE \cite{Stratonovich}).

The Black-Sholes PDE (\ref{BS}) can be applied to a number of
one-dimensional models of interpretations of prices given to $u$, e.g., puts
or calls, and to $s$, e.g., stocks or futures, dividends, etc. The most
important examples are European call and put options, defined by:
\begin{eqnarray}
&&u_{\mathrm{Call}}(s,t)=s\,\mathcal{N}(\mathrm{d_{1}})\,\mathrm{e}%
^{-T\delta }-k\,\mathcal{N}(\mathrm{d_{2}})\,\mathrm{e}^{-rT},  \label{call}
\\
&&u_{\mathrm{Put}}(s,t)=k\,\mathcal{N}(-\mathrm{d_{2}})\,\mathrm{e}^{-rT}-s\,%
\mathcal{N}(-\mathrm{d_{1}})\,\mathrm{e}^{-T\delta },  \label{put} \\
&&\,\mathcal{N}(\lambda )=\frac{1}{2}\left( 1+\mathrm{erf}\left( \frac{%
\lambda }{\sqrt{2}}\right) \right) ,  \notag \\
&&\mathrm{d_{1}}=\frac{\ln \left( \frac{s}{k}\right) +T\left( r-\delta +%
\frac{\sigma ^{2}}{2}\right) }{\sigma \sqrt{T}},\qquad \mathrm{d_{2}}=\frac{%
\ln \left( \frac{s}{k}\right) +T\left( r-\delta -\frac{\sigma ^{2}}{2}%
\right) }{\sigma \sqrt{T}},  \notag
\end{eqnarray}%
where erf$(\lambda )$ is the (real-valued) error function, $k$ denotes the
strike price and $\delta $ represents the dividend yield. In addition, for
each of the call and put options, there are five Greeks (see, e.g. \cite{Kelly,IvCogComp}), or sensitivities, which are partial derivatives of the
option-price with respect to stock price (Delta), interest rate (Rho),
volatility (Vega), elapsed time since entering into the option (Theta), and
the second partial derivative of the option-price with respect to the stock
price (Gamma).
\begin{figure}[htb]
\centering \includegraphics[width=10cm]{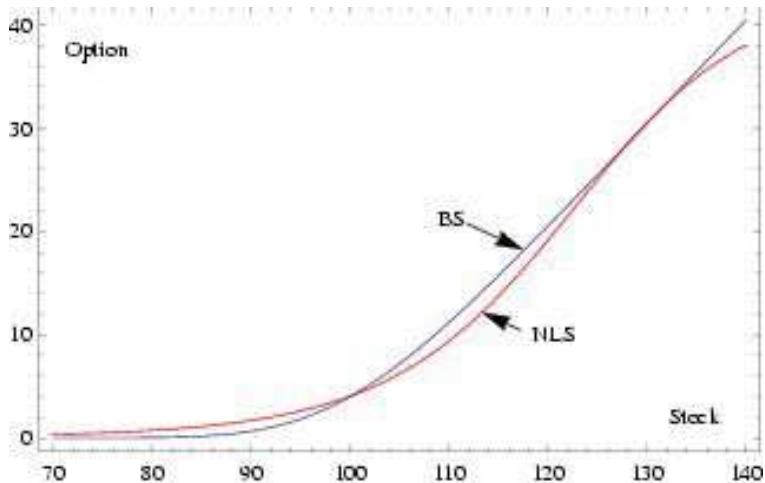}
\caption{Fitting the Black--Scholes call option with $\protect\beta (w)$%
-adaptive PDF of the shock-wave NLS-solution (\protect\ref{tanh1}).}
\label{fitCall}
\end{figure}
\begin{figure}[htb]
\centering \includegraphics[width=10cm]{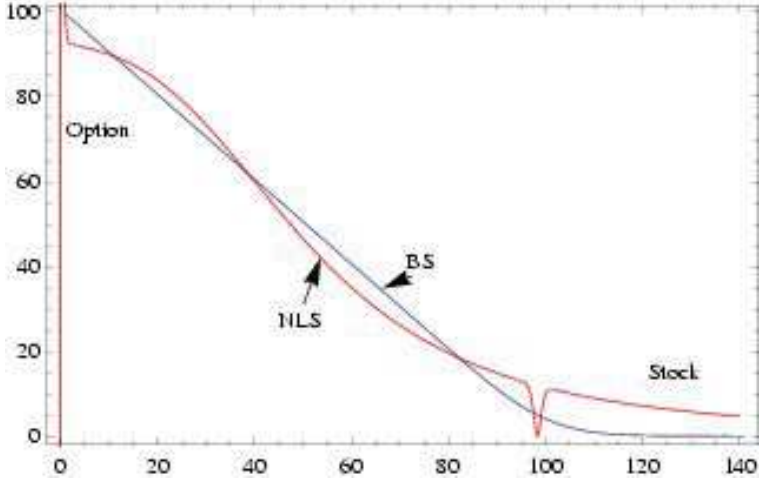}
\caption{Fitting the Black--Scholes put option with $\protect\beta (w)$%
--adaptive PDF of the shock-wave NLS $\protect\psi _{2}(s,t)$ solution (%
\protect\ref{tanh1}). Notice the kink near $s=100$.}
\label{fitPut}
\end{figure}

Using the standard \emph{Kolmogorov probability} approach, instead of the
market value of an option given by the Black--Scholes equation (\ref{BS}),
we could consider the corresponding probability density function (PDF) given
by the backward Fokker--Planck equation (see \cite{Gardiner,Voit}).
Alternatively, we can obtain the same PDF (for the market value of a stock
option), using the \emph{quantum--probability} formalism \cite%
{ComplexDyn,QuLeap}, as a solution to a time--dependent linear or nonlinear \emph{Schr\"{o}dinger equation} for the evolution of the complex--valued wave $\psi -$%
function for which the absolute square, $|\psi |^{2},$ is the PDF. The adaptive nonlinear Schr\"{o}dinger (NLS)
equation was recently used in \cite{IvCogComp} as an approach to option price modelling, as briefly reviewed in this section. The new model, philosophically founded on adaptive markets hypothesis \cite{Lo1,Lo2} and Elliott wave market theory \cite{Elliott1,Elliott2}, as well as my own recent work on quantum congition \cite{LSF,QnnBk},
describes adaptively controlled Brownian market behavior. This nonlinear approach to option price modelling is reviewed in the next section. Its important limiting case with low interest-rate reduces to the linear Schr\"{o}dinger equation. This linear approach to option price modelling is elaborated in the subsequent section.

\section{Nonlinear adaptive wave model for general option pricing}

\subsection{Adaptive NLS model}

The adaptive, wave--form, nonlinear and stochastic option--pricing model
with stock price $s,$ volatility $\sigma $ and interest rate $r$ is formally
defined as a complex-valued, focusing (1+1)--NLS equation, defining the
time-dependent \emph{option--price wave function} $\psi =\psi (s,t)$, whose
absolute square\ $|\psi (s,t)|^{2}$ represents the probability density
function (PDF) for the option price in terms of the stock price and time. In
natural quantum units, this NLS equation reads:
\begin{equation}
\mathrm{i}\partial _{t}\psi =-\frac{1}{2}\sigma \partial _{ss}\psi -\beta
|\psi |^{2}\psi ,\qquad (\mathrm{i}=\sqrt{-1}),\qquad   \label{nlsGen}
\end{equation}
where $\beta=\beta (r,w) $ denotes the adaptive market-heat potential (see \cite{KleinertBk}), so the term $V(\psi )=-\beta |\psi |^{2}$ represents the $%
\psi -$dependent potential field. In the simplest
nonadaptive scenario $\beta $ is equal to the interest rate $r$, while in
the adaptive case it depends on the set of adjustable synaptic weights $\{w^i_j\}$ as:
\begin{equation}
\beta (r,w)=r\sum_{i=1}^{n}w_{1}^{i}\,\text{erf}\left( \frac{w_{2}^{i}s}{%
w_{3}^{i}}\right) .  \label{betaW}
\end{equation}
Physically, the NLS equation (\ref{nlsGen}) describes a
nonlinear wave (e.g. in Bose-Einstein condensates) defined by the complex-valued wave function $\psi
(s,t)$ of real space and time parameters. In the present context, the
space-like variable $s$ denotes the stock (asset) price.

The NLS equation (\ref{nlsGen}) has been exactly solved using the power series
expansion method \cite{LiuEtAl01,LiuFan05} of {Jacobi elliptic functions}
\cite{AbrSte}. Consider the $\psi -$function describing a single plane wave, with the wave number $k$ and circular frequency $%
\omega $:
\begin{equation}
\psi (s,t)=\phi (\xi )\,\mathrm{e}^{\mathrm{i}(ks-\omega t)},\qquad \text{%
with \ }\xi =s-\sigma kt\text{ \ and \ }\phi (\xi )\in \Bbb{R}.
\label{subGen}
\end{equation}
Its substitution into the NLS equation (\ref{nlsGen}) gives the nonlinear
oscillator ODE:
\begin{equation}
\phi ^{\prime \prime }(\xi )+[\omega -\frac{1}{2}\sigma k^{2}]\,\phi (\xi
)+\beta \phi ^{3}(\xi )=0.  \label{15e}
\end{equation}

\bigskip We can seek a solution $\phi (\xi )$ for (\ref{15e}) as a linear function \cite{LiuFan05}
\[
\phi (\xi )=a_{0}+a_{1}\mathrm{sn}(\xi ),
\]
where $\mathrm{sn}(s)=\mathrm{sn}(s,m)$ are Jacobi elliptic sine functions
with \textit{elliptic modulus} $m\in \lbrack 0,1]$, such that $\mathrm{sn}%
(s,0)=\sin (s)\ $and $\mathrm{sn}(s,1)=\mathrm{\tanh }(s)$. The solution of (%
\ref{15e}) was calculated in \cite{IvCogComp} to be
\begin{eqnarray*}
\phi (\xi ) &=&\pm m\sqrt{\frac{-\sigma }{\beta }}\,\mathrm{sn}(\xi ),\qquad
~\text{for~~}m\in \lbrack 0,1];~~\text{and} \\
\phi (\xi ) &=&\pm \sqrt{\frac{-\sigma }{\beta }}\,\mathrm{\tanh }(\xi
),\qquad \text{for~~}m=1.
\end{eqnarray*}
This gives the exact periodic solution of (\ref{nlsGen}) as \cite{IvCogComp}
\begin{eqnarray}
\psi _{1}(s,t) &=&\pm m\sqrt{\frac{-\sigma }{\beta (w)}}\,\mathrm{sn}%
(s-\sigma kt)\,\mathrm{e}^{\mathrm{i}[ks-\frac{1}{2}\sigma
t(1+m^{2}+k^{2})]},\qquad ~~\text{for~~}m\in \lbrack 0,1);  \label{sn1} \\
\psi _{2}(s,t) &=&\pm \sqrt{\frac{-\sigma }{\beta (w)}}\,\mathrm{\tanh }%
(s-\sigma kt)\,\mathrm{e}^{\mathrm{i}[ks-\frac{1}{2}\sigma
t(2+k^{2})]},\qquad \qquad \text{for~~}m=1,  \label{tanh1}
\end{eqnarray}
where (\ref{sn1}) defines the general solution, while (\ref{tanh1}) defines
the \emph{envelope shock-wave}\footnote{%
A shock wave is a type of fast-propagating nonlinear disturbance that
carries energy and can propagate through a medium (or, field). It is
characterized by an abrupt, nearly discontinuous change in the
characteristics of the medium. The energy of a shock wave dissipates
relatively quickly with distance and its entropy increases. On the other
hand, a soliton is a self-reinforcing nonlinear solitary wave packet that
maintains its shape while it travels at constant speed. It is caused by a
cancelation of nonlinear and dispersive effects in the medium (or, field).}
(or, `dark soliton') solution of the NLS equation (\ref{nlsGen}).

Alternatively, if we seek a solution $\phi (\xi )$ as a linear function of
Jacobi elliptic cosine functions, such that $\mathrm{cn}(s,0)=\cos (s)$ and $%
\mathrm{cn}(s,1)=\mathrm{sech}(s)$,\footnote{%
A closely related solution of an anharmonic oscillator ODE:
\[
\phi ^{\prime \prime }(s)+\phi (s)+\phi ^{3}(s)=0
\]
is given by
\[
\phi (s)=\sqrt{\frac{2m}{1-2m}}\,\text{cn}\left( \sqrt{1+\frac{2m}{1-2m}}%
~s,\,m\right) .
\]
}
\[
\phi (\xi )=a_{0}+a_{1}\mathrm{cn}(\xi ),
\]
then we get \cite{IvCogComp}
\begin{eqnarray}
\psi _{3}(s,t) &=&\pm m\sqrt{\frac{\sigma }{\beta (w)}}\,\mathrm{cn}%
(s-\sigma kt)\,\mathrm{e}^{\mathrm{i}[ks-\frac{1}{2}\sigma
t(1-2m^{2}+k^{2})]},\qquad \text{for~~}m\in \lbrack 0,1);  \label{cn1} \\
\psi _{4}(s,t) &=&\pm \sqrt{\frac{\sigma }{\beta (w)}}\,\mathrm{sech}%
(s-\sigma kt)\,\mathrm{e}^{\mathrm{i}[ks-\frac{1}{2}\sigma
t(k^{2}-1)]},\qquad \qquad \text{for~~}m=1,  \label{sech1}
\end{eqnarray}
where (\ref{cn1}) defines the general solution, while (\ref{sech1}) defines
the \emph{envelope solitary-wave} (or, `bright soliton') solution of the NLS
equation (\ref{nlsGen}).

In all four solution expressions (\ref{sn1}), (\ref{tanh1}), (\ref{cn1}) and
(\ref{sech1}), the adaptive potential $\beta (w)$ is yet to be calculated
using either unsupervised Hebbian learning, or supervised
Levenberg--Marquardt algorithm (see, e.g. \cite{NeuFuz,CompMind}). In this
way, the NLS equation (\ref{nlsGen}) becomes the \emph{quantum neural network} (see
\cite{QnnBk}). Any kind of numerical analysis can be easily performed using
above closed-form solutions $\psi _{i}(s,t)~~(i=1,...,4)$ as initial
conditions.

The adaptive NLS--PDFs of the shock-wave type (\ref{tanh1}) has been used in \cite{IvCogComp} to
fit the Black--Scholes call and put options (see Figures \ref{fitCall} and \ref{fitPut}). Specifically, the adaptive heat potential (\ref{betaW}) was combined with the
spatial part of (\ref{tanh1})
\begin{equation}
\phi (s)=\left| \sqrt{\frac{\sigma }{\beta }}\tanh (s-kt\sigma )\right|
{}^{2},  \label{kink2}
\end{equation}
while parameter estimates where obtained using 100 iterations of the
Levenberg--Marquardt algorithm.
\begin{figure}[tbh]
\centering \includegraphics[width=10cm]{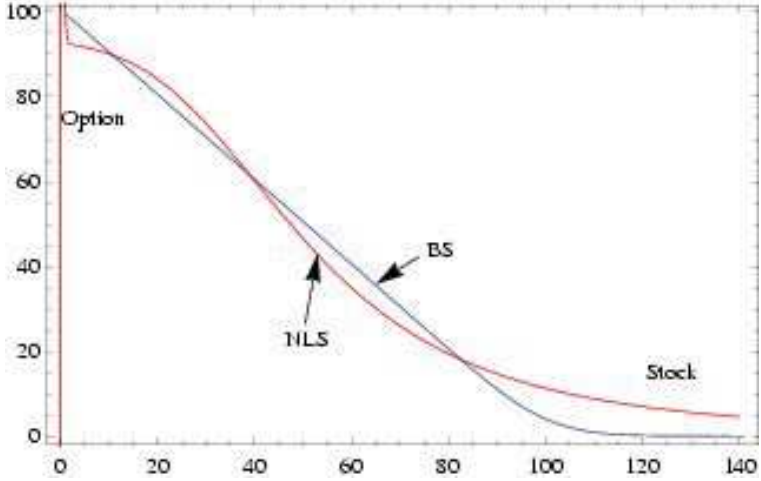}
\caption{Smoothing out the kink in the put option fit, by combining the
shock-wave solution with the soliton solution, as defined by (\ref{kinksech}%
).}
\label{tanSechPut}
\end{figure}

As can be seen from Figure (\ref{fitPut}) there is a kink near $s=100$. This
kink, which is a natural characteristic of the spatial shock-wave (\ref
{kink2}), can be smoothed out (Figure \ref{tanSechPut}) by taking the sum of the spatial parts of the
shock-wave solution (\ref{tanh1}) and the soliton solution (\ref
{sech1}) as:
\begin{equation}
\phi (s)=\left\vert \sqrt{\frac{\sigma }{\beta }}\left[d_{1}\tanh
(s-kt\sigma )+d_{2}\,\text{sech}(s-kt\sigma )\right] \right\vert {}^{2}.
\label{kinksech}
\end{equation}

The adaptive NLS--based Greeks (Delta, Rho, Vega, Theta and Gamma) have been
defined in \cite{IvCogComp}, as partial derivatives of the shock-wave
solution (\ref{tanh1}).

\newpage
\subsection{Adaptive Manakov system}

Next, for the purpose of including a \emph{controlled stochastic volatility}%
\footnote{%
Controlled stochastic volatility here represents volatility evolving in a
stochastic manner but within the controlled boundaries.} into the
adaptive--NLS model (\ref{nlsGen}), the full bidirectional quantum neural
computation model \cite{QnnBk} for option-price forecasting has been formulated in \cite
{IvCogComp} as a self-organized system of two coupled self-focusing NLS
equations: one defining the \emph{option--price wave function} $\psi =\psi
(s,t)$ and the other defining the \emph{volatility wave function} $\sigma
=\sigma (s,t)$:
\begin{eqnarray}
\text{Volatility NLS :}\quad \mathrm{i}\partial _{t}\sigma  &=&-\frac{1}{2}%
\partial _{ss}\mathcal{\sigma }-\beta (r,w)\left( |\mathcal{\sigma }%
|^{2}+|\psi |^{2}\right) \mathcal{\sigma },  \label{stochVol} \\
\text{Option price NLS :}\quad \mathrm{i}\partial _{t}\psi  &=&-\frac{1}{2}%
\partial _{ss}\psi -\beta (r,w)\left( |\mathcal{\sigma }|^{2}+|\psi
|^{2}\right) \psi .  \label{stochPrice}
\end{eqnarray}
In this coupled model, the $\sigma $--NLS (\ref{stochVol}) governs the $%
(s,t)-$evolution of stochastic volatility, which plays the role of a
nonlinear coefficient in (\ref{stochPrice}); the $\psi $--NLS (\ref
{stochPrice}) defines the $(s,t)-$evolution of option price, which plays the
role of a nonlinear coefficient in (\ref{stochVol}). The purpose of this
coupling is to generate a \emph{leverage effect}, i.e. stock volatility is
(negatively) correlated to stock returns\footnote{The hypothesis that financial
leverage can explain the leverage effect was
first discussed by F. Black \cite{Bl76}.} (see, e.g. \cite{Roman}). This
bidirectional associative memory effectively performs quantum neural
computation \cite{QnnBk}, by giving a spatio-temporal and quantum
generalization of Kosko's BAM family of neural networks \cite{Kosko1,Kosko2}%
. In addition, the shock-wave and solitary-wave nature of the coupled NLS
equations may describe brain-like effects frequently occurring in financial
markets: volatility/price propagation, reflection and collision of shock and
solitary waves (see \cite{Han}).

\begin{figure}[tbh]
\centering \includegraphics[width=12cm]{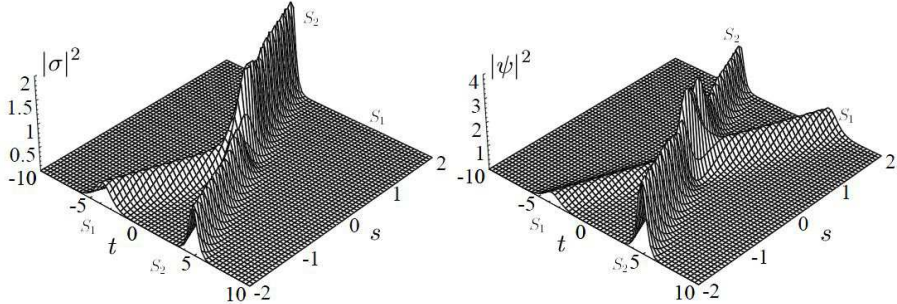}
\caption{Hypothetical market scenario including sample PDFs for volatility $|%
\mathcal{\protect\sigma }|^{2}$ and $|\protect\psi |^{2}$ of the Manakov
2--soliton (\ref{ManSol}). On the left, we observe the $(s,t)-$evolution of
stochastic volatility: we have a collision of two volatility
component-solitons, $S_{1}(s,t)$ and $S_{2}(s,t)$, which join together into
the resulting soliton $S_{2}(s,t)$, annihilating the $S_{1}(s,t)$ component
in the process. On the right, we observe the $(s,t)-$evolution of option
price: we have a collision of two option component-solitons, $S_{1}(s,t)$
and $S_{2}(s,t)$, which pass through each other without much change, except
at the collision point. Due to symmetry of the Manakov system, volatility
and option price can exchange their roles.}
\label{SolitonCollision}
\end{figure}

The coupled NLS-system (\ref{stochVol})--(\ref{stochPrice}), without an
embedded $w-$learning (i.e., for constant $\beta =r$ -- the interest rate),
actually defines the well-known \emph{Manakov system},\footnote{%
Manakov system has been used to describe the interaction between wave
packets in dispersive conservative media, and also the interaction between
orthogonally polarized components in nonlinear optical fibres (see, e.g.
\cite{Kerr,Yang1} and references therein).} proven by S. Manakov in 1973
\cite{manak74} to be completely integrable, by the existence of infinite
number of involutive integrals of motion. It admits `bright' and `dark'
soliton solutions. The simplest solution of (\ref{stochVol})--(\ref
{stochPrice}), the so-called \textit{Manakov bright 2--soliton}, has the
form resembling that of the sech-solution (\ref{sech1}) (see \cite
{Benney,Zakharov,Hasegawa,Radhakrishnan,Agrawal,Yang,Elgin}), and is formally defined by:
\begin{equation}
\mathbf{\psi }_{\mathrm{sol}}(s,t)=2b\,\mathbf{c\,}\mathrm{sech}(2b(s+4at))\,%
\mathrm{e}^{-2\mathrm{i}(2a^{2}t+as-2b^{2}t)},  \label{ManSol}
\end{equation}
where $\mathbf{\psi }_{\mathrm{sol}}(s,t)=\left(
\begin{array}{c}
\sigma (s,t) \\
\psi (s,t)
\end{array}
\right) $, $\mathbf{c}=(c_{1},c_{2})^{T}$ is a unit vector such that $%
|c_{1}|^{2}+|c_{2}|^{2}=1$. Real-valued parameters $a$ and $b$ are some
simple functions of $(\sigma ,\beta ,k)$, which can be determined by the Levenberg--Marquardt algorithm. I have argued in \cite{IvCogComp} that in some
short-time financial situations, the adaptation effect on $\beta$ can be
neglected, so our option-pricing model (\ref{stochVol})--(\ref{stochPrice})
can be reduced to the Manakov 2--soliton model (\ref{ManSol}), as depicted
and explained in Figure \ref{SolitonCollision}.

\newpage

\section{Quantum wave model for low interest-rate
option pricing}

In the case of a low interest-rate $r\ll 1$, we have $\beta (r)\ll 1$, so $%
V(\psi )\rightarrow 0,$ and therefore equation (\ref{nlsGen}) can be
approximated by a quantum-like \emph{option wave packet.} It is defined by a
continuous superposition of \emph{de Broglie's plane waves}, `physically'
associated with a free quantum particle of unit mass. This linear wave
packet, given by the time-dependent complex-valued wave function $\psi =\psi
(s,t)$, is a solution of the \emph{linear Schr\"{o}dinger equation} with
zero potential energy, Hamiltonian operator $\hat{H}$ and volatility $\sigma
$ playing the role similar to the Planck constant. This equation can be
written as:
\begin{equation}
\mathrm{i}\sigma \partial _{t}\psi =\hat{H}\psi ,\qquad \text{where}\qquad
\hat{H}=-\frac{\sigma ^{2}}{2}\partial _{ss}.  \label{sch1}
\end{equation}

Thus, we consider the $\psi -$function describing a single de Broglie's
plane wave, with the wave number $k$, linear momentum $p=\sigma k,$
wavelength $\lambda _{k}=2\pi /k,$\ angular frequency $\omega _{k}=\sigma
k^{2}/2,$ and oscillation period $T_{k}=2\pi /\omega _{k}=4\pi /\sigma k^{2}$%
. It is defined by (compare with \cite{Griffiths,Thaller,QuLeap})
\begin{equation}
\psi _{k}(s,t)=A\mathrm{e}^{\mathrm{i}(ks-\omega _{k}t)}=A\mathrm{e}^{%
\mathrm{i}(ks-{\frac{\sigma k^{2}}{2}}t)}=A\cos (ks-{\frac{\sigma k^{2}}{2}}%
t)+A\mathrm{i}\sin (ks-{\frac{\sigma k^{2}}{2}}t),  \label{Broglie}
\end{equation}
where $A$ is the amplitude of the wave, the angle $(ks-\omega _{k}t)=(ks-{%
\frac{\sigma k^{2}}{2}}t)$\ represents the phase of the wave $\psi _{k}$
with the \emph{phase velocity:} $v_{k}=\omega _{k}/k=\sigma k/2.$

The space-time wave function $\psi (s,t)$ that satisfies the linear
Schr\"{o}dinger equation (\ref{sch1}) can be decomposed (using Fourier's
separation of variables) into the spatial part $\phi (s)\,$\ and the
temporal part $\mathrm{e}^{-\mathrm{i}\omega t}\ $as:
\[
\psi (s,t)=\phi (s)\,\mathrm{e}^{-\mathrm{i}\omega t}=\phi (s)\,\mathrm{e}^{-%
\frac{\mathrm{i}}{\sigma }Et}.
\]
The spatial part, representing \emph{stationary }(or,\emph{\ amplitude})%
\emph{\ wave function}, $\phi (s)=A\mathrm{e}^{\mathrm{i}ks},$ satisfies the
\emph{linear harmonic oscillator,} which can be formulated in several
equivalent forms:
\begin{equation}
\phi ^{\prime\prime}+k^{2}\phi =0,\qquad \phi ^{\prime\prime}+\left( \frac{p%
}{\sigma }\right) ^{2}\phi =0,\qquad \phi ^{\prime\prime}+\left( \frac{%
\omega _{k}}{v_{k}}\right) ^{2}\phi =0,\qquad \phi ^{\prime\prime}+\frac{%
2E_{k}}{\sigma ^{2}}\phi =0.  \label{stac}
\end{equation}

Planck's \emph{energy quantum} of the option wave $\psi _{k}$ is given by: $
E_{k}=\sigma \omega _{k}=\frac{1}{2}(\sigma k)^{2}.
$

From the plane-wave expressions (\ref{Broglie}) we have: $\psi _{k}(s,t)=A%
\mathrm{e}^{\frac{\mathrm{i}}{\sigma }(ps-E_{k}t)}-$ for the wave going to
the `right' and $\psi _{k}(s,t)=A\mathrm{e}^{-\frac{\mathrm{i}}{\sigma }%
(ps+E_{k}t)}-$ for the wave going to the `left'.

The general solution to (\ref{sch1}) is formulated as a linear combination
of de Broglie's option waves (\ref{Broglie}), comprising the option wave-packet:
\begin{equation}
\psi (s,t)=\sum_{i=0}^{n}c_{i}\psi _{k_{i}}(s,t),\qquad (\text{with}\ n\in
\Bbb{N}).  \label{w-pack}
\end{equation}
Its absolute square, $|\psi (s,t)|^{2},$ represents the probability density
function at a time $t.$

The \emph{group velocity} of an option wave-packet is given by: $\ v_{g}=d\omega
_{k}/dk.$ It is related to the phase velocity $v_{k}$ of a plane wave as: $%
v_{g}=v_{k}-\lambda _{k}dv_{k}/d\lambda _{k}.$ Closely related is the \emph{center} of the option wave-packet (the point of maximum amplitude), given by: $%
s=td\omega _{k}/dk.$\newline

The following quantum-motivated assertions can be stated:

\begin{enumerate}
\item  Volatility $\sigma $ has dimension of \emph{financial action}, or
\emph{energy }$\times $\emph{\ time}.

\item  The total energy $E$ of an option wave-packet is (in the case of similar plane waves) given by Planck's
superposition of the energies $E_{k}$ of $n$ individual waves:  $E=n\sigma \omega
_{k}=\frac{n}{2}(\sigma k)^{2},$ where $L=n\sigma $ denotes the \emph{angular
momentum} of the option wave-packet, representing the shift between its growth and decay, and \emph{vice versa.}

\item  The average energy $\left\langle E\right\rangle $\ of an option wave-packet
is given by Boltzmann's partition function:
\[
\left\langle E\right\rangle =\frac{\sum_{n=0}^{\infty }nE_{k}\mathrm{e}^{-%
\frac{nE_{k}}{bT}}}{\sum_{n=0}^{\infty }\mathrm{e}^{-\frac{nE_{k}}{bT}}}=%
\frac{E_{k}}{\mathrm{e}^{\frac{E_{k}}{bT}}-1},
\]
where $b$ is the Boltzmann-like kinetic constant and $T$ is the market temperature.

\item  The energy form of the Schr\"{o}dinger equation (\ref{sch1}) reads: $%
E\psi =\mathrm{i}\sigma \partial _{t}\psi $.

\item  The eigenvalue equation for the Hamiltonian operator $\hat{H}$ is the
\emph{stationary Schr\"{o}dinger equation:} $\ $%
\[
\hat{H}\phi (s)=E\phi (s),\qquad \text{or}\qquad E\phi (s)=-\frac{\sigma ^{2}%
}{2}\partial _{ss}\phi (s),
\]
which is just another form of the harmonic oscillator (\ref{stac}). It has
oscillatory solutions of the form:
\[
\phi _{E}(s)=c_{1}\mathrm{e}^{\frac{\mathrm{i}}{\sigma }\sqrt{2E_{k}}%
\,s}+c_{2}\mathrm{e}^{-\frac{\mathrm{i}}{\sigma }\sqrt{2E_{k}}\,s}\,,
\]
called \emph{energy eigen-states} with energies $E_{k}$ and denoted by: $\hat{H}%
\phi _{E}(s)=E_{k}\phi _{E}(s).$
\end{enumerate}
\begin{figure}[tbh]
\centering \includegraphics[width=10cm]{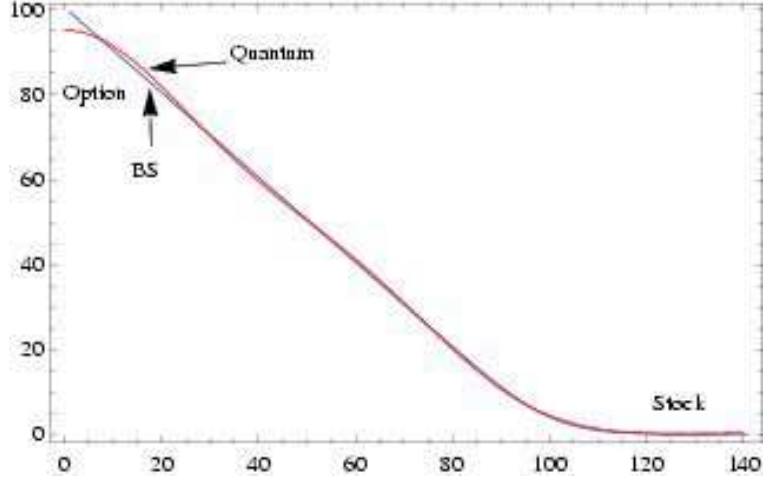}
\caption{Fitting the Black--Scholes put option with the quantum PDF
given by the absolute square of (\ref{w-pack}) with $n=7$.}
\label{PutQuant}
\end{figure}
\begin{figure}[tbh]
\centering \includegraphics[width=10cm]{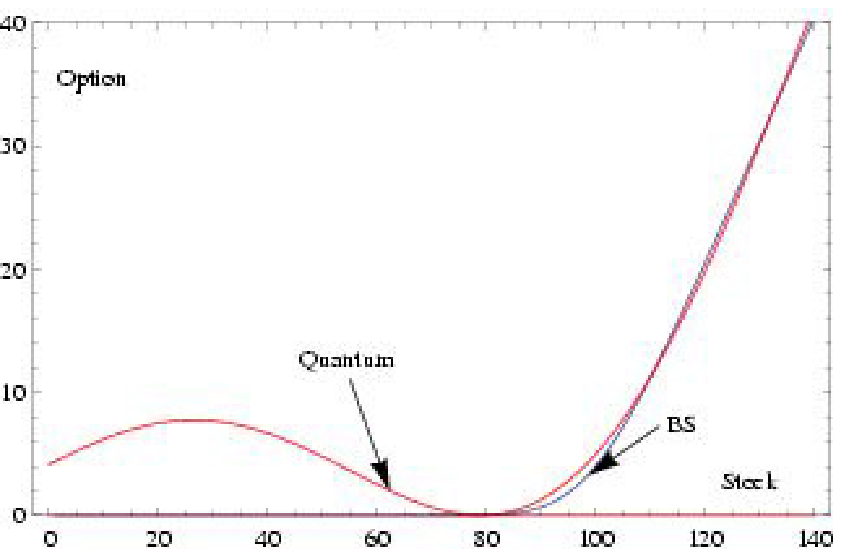}
\caption{Fitting the Black--Scholes call option with the quantum PDF
given by the absolute square of (\ref{w-pack}) with $n=3$. Note that fit is good in the realistic stock region: $s\in [75,140]$.}
\label{QuantumCall}
\end{figure}

The Black--Scholes put and call options have been fitted with the quantum
PDFs (see Figures \ref{PutQuant} and \ref{QuantumCall}) given by
the absolute square of (\ref{w-pack}) with $n=7$ and $n=3$, respectively. Using supervised Levenberg--Marquardt algorithm and \emph{Mathematica} 7, the following coefficients were obtained for the Black--Scholes put option:\newline

$\sigma^* = -0.0031891,~t^*= -0.0031891,~k_1=
   2.62771,~k_2= 2.62777,~k_3= 2.65402,$

   $k_4=
   2.61118,~k_5= 2.64104,~k_6= 2.54737,~k_7=
   2.62778,~c_1= 1.26632,~c_2= 1.26517,$

   $c_3=
   2.74379,~c_4= 1.35495,~c_5= 1.59586,~c_6=
   0.263832,~c_7= 1.26779,$

   with ~$\sigma_{BS}=-94.0705\sigma^*,~t_{BS}=-31.3568t^*.$\newline

\noindent Using the same algorithm, the following coefficients were obtained for the Black--Scholes call option:\newline

$\sigma^* = -11.9245,~t^*=
   -11.9245,~k_1= 0.851858,~k_2=
   0.832409,$

$k_3= 0.872061,~c_1=
   2.9004,~c_2= 2.72592,~c_3=
   2.93291,$

   with  $\sigma_{BS}-0.0251583 \sigma^*,~t=-0.00838609 t^*.$\newline

Now, given some initial option wave function, $\psi (s,0)=\psi _{0}(s),$ a solution
to the initial-value problem for the linear Schr\"{o}dinger equation (\ref
{sch1}) is, in terms of the pair of Fourier transforms $(\mathcal{F},%
\mathcal{F}^{-1}),$ given by (see \cite{Thaller})
\begin{equation}
\psi (s,t)=\mathcal{F}^{-1}\left[ \mathrm{e}^{-\mathrm{i}\omega t}\mathcal{F}%
(\psi _{0})\right] =\mathcal{F}^{-1}\left[ \mathrm{e}^{-\mathrm{i}{\frac{%
\sigma k^{2}}{2}}t}\mathcal{F}(\psi _{0})\right] .  \label{Fouri}
\end{equation}

For example (see \cite{Thaller}), suppose we have an initial option wave-function
at time $t=0$ given by the complex-valued Gaussian function:
\[
\psi (s,0)=\mathrm{e}^{-as^{2}/2}\mathrm{e}^{\mathrm{i}\sigma ks},
\]
where $a$ is the width of the Gaussian, while $p$ is the average momentum of
the wave. Its Fourier transform, $\hat{\psi}_{0}(k)=\mathcal{F}[\psi (s,0)],$
is given by
\[
\hat{\psi}_{0}(k)=\frac{\mathrm{e}^{-\frac{(k-p)^{2}}{2a}}}{\sqrt{a}}.
\]
The solution at time $t$ of the initial value problem is given by
\[
\psi (s,t)=\frac{1}{\sqrt{2\pi a}}\int_{-\infty }^{+\infty }\mathrm{e}^{%
\mathrm{i}(ks-{\frac{\sigma k^{2}}{2}}t)}\,\mathrm{e}^{-\frac{a(k-p)^{2}}{2a}%
}\,dk,
\]
which, after some algebra becomes
\[
\psi (s,t)=\frac{\mathrm{\exp }(-\frac{as^{2}-2\mathrm{i}sp+\mathrm{i}p^{2}t%
}{2(1+\mathrm{i}at)})}{\sqrt{1+\mathrm{i}at}},\qquad (\text{with \ }p=\sigma
k).
\]

As a simpler example,\footnote{An example of a more general Gaussian wave-packet solution of (\ref{sch1})
is given by:
\[
\psi (s,t)=\sqrt{\frac{\sqrt{a/\pi }}{1+\mathrm{i}at}}\,\exp \left( \frac{-%
\frac{1}{2}a(s-{s_{0}})^{2}-\frac{\mathrm{i}}{2}p_{0}^{2}t+\mathrm{i}p_{0}(s-%
{s_{0}})}{1+\mathrm{i}at}\right) ,
\]
where $s_{0},p_{0}$ are initial stock-price and average momentum, while $a$
is the width of the Gaussian. At time $t=0$ the `particle' is at rest around
$s=0$, its average momentum $p_{0}=0$. The wave function spreads with time
while its maximum decreases and stays put at the origin. At time $-t$ the
wave packet is the complex-conjugate of the wave-packet at time $t$.} if we have an initial option wave-function given by the
real-valued Gaussian function,
\[
\psi (s,0)=\frac{\mathrm{e}^{-s^{2}/2}}{\sqrt[4]{\pi }},
\]
the solution of (\ref{sch1}) is given by the complex-valued $\psi -$%
function,
\[
\psi (s,t)=\frac{\mathrm{\exp }(-\frac{s^{2}}{2(1+\mathrm{i}t)})}{\sqrt[4]{%
\pi }\sqrt{1+\mathrm{i}t}}.
\]

From (\ref{Fouri}) it follows that a stationary option wave-packet is given by:
\[
\phi (s)=\frac{1}{\sqrt{2\pi }}\int_{-\infty }^{+\infty }\mathrm{e}^{\frac{%
\mathrm{i}}{\sigma }ks}\,\hat{\psi}(k)\,dk,\qquad \text{where}\qquad \hat{%
\psi}(k)=\mathcal{F}[\phi (s)].
\]
As $|\phi (s)|^{2}$ is the stationary stock PDF, we can calculate the \emph{expectation values} of the
stock and the wave number of the whole option wave-packet, consisting of $n$ measured
plane waves, as:
\begin{equation}
\left\langle s\right\rangle =\int_{-\infty }^{+\infty }s|\phi
(s)|^{2}ds\qquad \text{and}\qquad \left\langle k\right\rangle =\int_{-\infty
}^{+\infty }k|\hat{\psi}(k)|^{2}dk.  \label{means}
\end{equation}
The recordings of $n$ individual option plane waves (\ref{Broglie}) will be
scattered around the mean values (\ref{means}). The width of the
distribution of the recorded $s-$ and $k-$values are uncertainties $\Delta s$
and $\Delta k,$ respectively. They satisfy the Heisenberg-type uncertainty
relation:
\[
\Delta s\,\Delta k\geq \frac{n}{2},
\]
which imply the similar relation for the total option energy and time:
\[
\Delta E\,\Delta t\geq \frac{n}{2}.
\]

Finally, Greeks for both put and call options are defined as the following partial derivatives of the option $\psi-$function PDF:\\

\noindent ${\text{Delta}}=\partial _{s}|\psi (s,t)|^{2}=\newline
{2\mathrm{i}\sum_{j=1}^{n}c_{j}k_{j}\,\mathrm{e}^{k_{j}\left( \mathrm{i}s-%
\text{${\frac{1}{2}}$}\mathrm{i}\sigma k_{j}t\right) }\text{Abs}\left[
\sum_{j=1}^{n}c_{j}\,\mathrm{e}^{k_{j}\left( \mathrm{i}s-\text{${\frac{1}{2}}
$}\mathrm{i}\sigma k_{j}t\right) }\right] \text{Abs}^{\prime }\left[
\sum_{j=1}^{n}c_{j}\,\mathrm{e}^{k_{j}\left( \mathrm{i}s-\text{${\frac{1}{2}}
$}\mathrm{i}\sigma k_{j}t\right) }\right] }$\\

\noindent ${\text{Vega}}=\partial _{{\text{$\sigma $}}}|\psi (s,t)|^{2}=%
\newline
{-\text{it}\sum_{j=1}^{n}c_{j}k_{j}{}^{2}\,\mathrm{e}^{k_{j}\left( \mathrm{i}%
s-\text{${\frac{1}{2}}$}\mathrm{i}\sigma k_{j}t\right) }\text{Abs}\left[
\sum_{j=1}^{n}c_{j}\,\mathrm{e}^{k_{j}\left( \mathrm{i}s-\text{${\frac{1}{2}}
$}\mathrm{i}\sigma k_{j}t\right) }\right] \text{Abs}^{\prime }\left[
\sum_{j=1}^{n}c_{j}\,\mathrm{e}^{k_{j}\left( \mathrm{i}s-\text{${\frac{1}{2}}
$}\mathrm{i}\sigma k_{j}t\right) }\right] }$\\

\noindent ${\text{Theta}}=\partial _{{\text{$t$}}}|\psi (s,t)|^{2}=\newline
{-\text{\textrm{i}$\sigma $}\sum_{j=1}^{n}c_{j}k_{j}{}^{2}\,\mathrm{e}%
^{k_{j}\left( \mathrm{i}s-\text{${\frac{1}{2}}$}\mathrm{i}\sigma
k_{j}t\right) }\text{Abs}\left[ \sum_{j=1}^{n}c_{j}\,\mathrm{e}^{k_{j}\left(
\mathrm{i}s-\text{${\frac{1}{2}}$}\mathrm{i}\sigma k_{j}t\right) }\right]
\text{Abs}^{\prime }\left[ \sum_{j=1}^{n}c_{j}\,\mathrm{e}^{k_{j}\left(
\mathrm{i}s-\text{${\frac{1}{2}}$}\mathrm{i}\sigma k_{j}t\right) }\right] }$\\

\noindent ${\text{Gamma}}=\partial _{ss}|\psi (s,t)|^{2}=\newline
{-2\sum_{j=1}^{n}c_{j}k_{j}{}^{2}\,\mathrm{e}^{k_{j}\left( \mathrm{i}s-\text{%
${\frac{1}{2}}$}\mathrm{i}\sigma k_{j}t\right) }\text{Abs}\left[
\sum_{j=1}^{n}c_{j}\,\mathrm{e}^{k_{j}\left( \mathrm{i}s-\text{${\frac{1}{2}}
$}\mathrm{i}\sigma k_{j}t\right) }\right] \text{Abs}^{\prime}\left[
\sum_{j=1}^{n}c_{j}\,\mathrm{e}^{k_{j}\left( \mathrm{i}s-\text{${\frac{1}{2}}
$}\mathrm{i}\sigma k_{j}t\right) }\right] -}\newline
{\left( \sum_{j=1}^{n}c_{j}k_{j}\,\mathrm{e}^{k_{j}\left( \mathrm{i}s-\text{$%
{\frac{1}{2}}$}\mathrm{i}\sigma k_{j}t\right) }\right) {}^{2}\text{Abs}%
^{\prime }\left[ \sum_{j=1}^{n}c_{j}\,\mathrm{e}^{k_{j}\left( \mathrm{i}s-%
\text{${\frac{1}{2}}$}\mathrm{i}\sigma k_{j}t\right) }\right] {}^{2}-}%
\newline
{\left( \sum_{j=1}^{n}c_{j}k_{j}\,\mathrm{e}^{k_{j}\left( \mathrm{i}s-\text{$%
{\frac{1}{2}}$}\mathrm{i}\sigma k_{j}t\right) }\right) {}^{2}\text{Abs}\left[
\sum_{j=1}^{n}c_{j}\,\mathrm{e}^{k_{j}\left( \mathrm{i}s-\text{${\frac{1}{2}}
$}\mathrm{i}\sigma k_{j}t\right) }\right] \text{Abs}^{\prime \prime }\left[
\sum_{j=1}^{n}c_{j}\,\mathrm{e}^{k_{j}\left( \mathrm{i}s-\text{${\frac{1}{2}}
$}\mathrm{i}\sigma k_{j}t\right) }\right] },$\\

\noindent where $\text{Abs}$ denotes the absolute value, while $\text{Abs}^{\prime}$ and $\text{Abs}^{\prime \prime }$ denote its first and second derivatives.

\bigbreak
\section{Conclusion}

I have proposed an adaptive--wave alternative to the standard
Black-Scholes option pricing model. The new model, philosophically founded
on adaptive markets hypothesis \cite{Lo1,Lo2} and Elliott wave market theory
\cite{Elliott1,Elliott2}, describes adaptively controlled Brownian market
behavior. Two approaches have been proposed: (i) a nonlinear one based on the adaptive NLS (solved by means of Jacobi
elliptic functions) and the adaptive Manakov system (of two coupled NLS equations); (ii) a linear quantum-mechanical one based on the free-particle Schr\"{o}dinger equation and de Broglie's plane waves. For the purpose of fitting the Black-Scholes data, the Levenberg-Marquardt algorithm was used.

The presented adaptive and quantum wave models are
spatio-temporal dynamical systems of much higher complexity \cite{ComNonlin} then the Black-Scholes model. This makes the
new wave models harder to analyze, but at the same time, their immense variety
is potentially much closer to the real financial market complexity,
especially at the time of financial crisis.

\end{document}